\documentclass{svjour2}
\smartqed 
\usepackage{latexsym}
\usepackage{graphicx}

\oddsidemargin=\evensidemargin

\newcommand{\beq}      {\begin{eqnarray}}
\newcommand{\eeq}      {\end{eqnarray}}
\newcommand{\rf}[1]    {(\ref{#1})}

\newcommand{\lp}       {\left}
\newcommand{\rp}       {\right}

\journalname{ }

\begin{document}

\title{Determination of density and concentration from fluorescent images of a gas flow %distribution in hypersonic underexpanded jets through fluorescent images.
%\thanks{Grants or other notes about the article that should go on the front page should be
%placed here. General acknowledgments should be placed at the end of the article.}
}
%\subtitle{Do you have a subtitle?\\ If so, write it here}
%\titlerunning{Short form of title}  % if too long for running head

\author{Marco Belan \and Sergio De Ponte \and Daniela Tordella}
%\authorrunning{Short form of author list} % if too long for running head

\institute{M. Belan \and S. De Ponte \at
Dipartimento di Ingegneria Aeronautica e Spaziale, Politecnico di Milano\\ 
%Tel.: +39-02-23998045\\
%Fax:  +39-02-23998028\\
%\email{belan@aero.polimi.it}           
%  \emph{Present address:} of F. Author  %  if needed
\and
D. Tordella (corresponding author) \at Dipartimento di Ingegneria Aeronautica e Spaziale, Politecnico di Torino\\ 
Tel.: +39-011-0296812\\
Fax:  +39-011-0296899\\
\email{daniela.tordella@polito.it}  }

\date{Received: date / Accepted: date}
%The correct dates will be entered by the editor

\maketitle

\begin{abstract}
%Insert your abstract here. Include keywords, PACS and mathematical
%subject classification numbers as needed.

A fluorescent image analysis procedure   to determine the distribution of species concentration and density in a gas flow is proposed.
The fluorescent emission is due to the excitation of atoms/molecules of a gas that is intercepted by an electron blade. 
The intensity of the fluorescent light is proportional to the local number density  of the gas. 
When the gas flow is a mixture of  different species, this proportionality can be  exploited to extract the contribution associated to the species from the spectral superposition acquired by a digital camera. This yields a method that simultaneously reveals species concentrations and   mass density of the mixture.
The procedure is applied to two under-expanded sonic jets discharged into a different gas ambient - Helium into Argon and Argon into Helium - to measure the concentration and density distribution along the jet axis and across it.  A comparison with experimental and numerical results obtained by other authors when observing under-expanded jets at different Mach numbers is made  with the density distribution along the axis of the jet. This density distribution appears to be self-similar.

\end{abstract}

\section{Introduction}

%%%%% literature
The density measurement technique in gas flows performed by means of an electron beam was  
described extensively by Muntz (1968) in an early review, that also reported the first attempts to obtain species concentration using filters in the optical system.

Electron beam techniques were also well described by B\"utefisch and Vennemann (1974). 
 
Later on, spectral measurements were performed on gas flows excited by an electron beam. These measurements yielded concentration values in isolated points in  space. 
Cattolica (1988) presented  experiments on the use of electron beam and laser induced fluorescence in high speed flows, where the concentration of nitric oxide was measured by means of spectroscopic methods.
%\cite{catt}
Another spectral study of the fluorescent emission of nitrogen, helium, and nitric oxide in a hypersonic wind tunnel along 
with an  analysis of background noise effects was reported by Price et al. (1992).
%\cite{price}
Recent reviews in this field,  which include the electron beam technique and species concentration measurement 
in hypersonic flows, have been written by Dankert et al. (1993) and  Gochberg (1994). The applicability of  relevant methods has also been discussed  for hot gases in high enthalpy flows, and at relatively high densities.
%\cite{dank} 
%\cite{goch} 
The extension of the electron-beam techniques to higher density-and-temperature test flows is possible thanks to the introduction of the pulsed electron beam - that replaces the continuous one -  which was first  obtained by Lutfy and Muntz (1999), 
%\cite{lutf} 
see also the recent work by Wehrmeyer (2006).

The electron-beam measurements that have been found in literature  are usually carried out in isolated spatial points. These measurements are usually very accurate for spectral scales, i.e. the gas emission is often analyzed for each  single spectrum line.  
Other techniques, such as  high-speed rainbow schlieren deflectometry (RSD), are more effective in reconstructing  flow spatial evolution as they allow instantaneous concentration contours to be determined across  the field  (Allamar et al. 1998, Agrawal et al. 2002, Yildirim and Agrawal, 2005). 
%\cite(yild)

The experimental technique  herein described aims at the simultaneous measurement of the distribution of the species
concentration and density in sections of a gas flow crossed by an electron-beam blade. 
The measurement is based on the analysis of fluorescent images obtained through   gas ionization induced by an
electron beam and acquired by a classic Charge Coupled Device (CCD) camera. The signal obtained from the color CCD
yields, within each pixel,  the projection of the fluorescence spectrum on three wavelength bands relevant to the standard
color Red, Green and Blue (RGB). By means of a suitable data post-processing, the data collected from a camera image are 
transformed into concentration and density maps. The aim is to obtain fast and simultaneous measurements on a large section of the test flow. The  technique is applied here to the measurement of axial and transversal  density and
concentration distributions in under-expanded hypersonic gas jets issued in a different gas ambient. The work is organized as follows.
The experimental equipment and the test flow are described in Section 2. Section 3 presents the data analysis algorithm.
Section 4 contains the results and  discussion. The concluding remarks are given in Section 5.

\section{Experimental set-up and test flow characteristics}

\begin{figure}[ht]
\centering
\includegraphics[width=12cm]{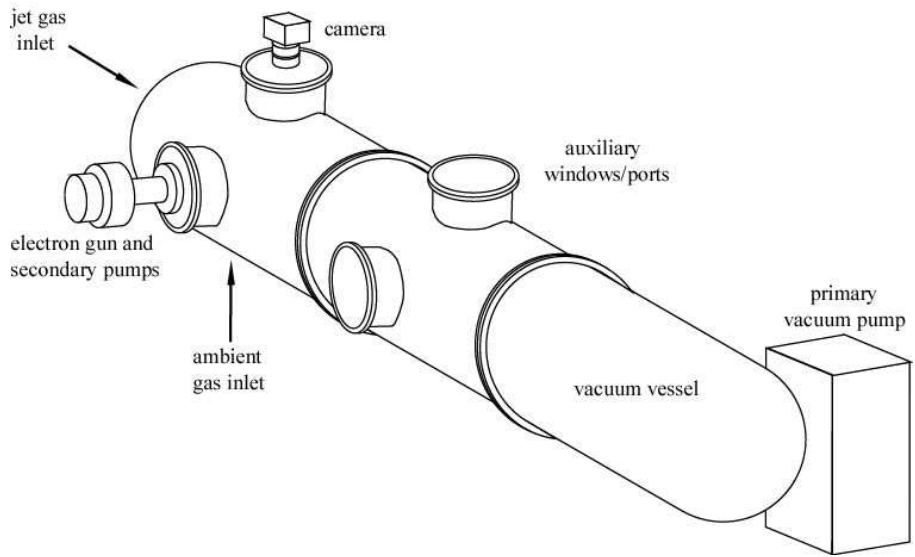}
\caption{Experimental apparatus. The vacuum vessel is shown in the 3-section configuration, two extra sections are available.}
\label{vessel}
\end{figure}

The study and implementation of the present technique are carried out in an apparatus designed for the study of hypersonic jets (Belan et al., 2001, a-b).
%\cite{bel1,bel2}
In this paper, we consider underexpanded jets obtained from truncated sonic nozzles and flowing along the longitudinal axis of a cylindrical vacuum vessel. The vessel is modular, with a total length of 5m and a diameter of 0.5m.  The  diameter of the orifice of the nozzles ranges from 0.3 to 2mm,  the vessel diameter is thus much larger than the diameter of the jets. As a consequence, the wall effects are limited. 
The apparatus is shown in fig. \ref{vessel}, where the 3 module configuration is represented (up to 5 sections can be mounted).
The vessel is equipped with a system of valves for the control of both the jet issued by the nozzles and the ambient gas. An electron gun operating
at very low pressures, thanks to a set of secondary pumps, and a color CCD camera can be mounted onto several ports and optical windows.

The stagnation pressure $p_0$ of the jets (at $T=300K$) can be varied in the range 2000 to $2 \cdot 10^5$Pa. 
The pressure in the vessel can be adjusted from 1.5 to 200 Pa by varying the volume flow of the primary pump. The resulting stagnation/ambient pressure ratio, $p_0/p_{\rm amb}$, ranges over 5 orders in magnitude. The stagnation pressures are measured by 3-digit resolution instruments, whilst the pressures in the vessel are monitored by means of 4-digit transducers. 

In this experiment, it is possible to use different gases for the jet and the surrounding ambient in the vessel. This makes it possible to vary the density ratio $\rho_{\rm jet} / \rho_{\rm amb}$  (where $\rho_{\rm jet}$ is the long term jet density far from the nozzle) as an independent parameter. The density ratio in the far field can range from 0.04 to 45 for 
jet/ambient gas  couples chosen from  air, helium, argon and xenon. 

The jets are characterized by the presence of a barrel shock and a normal shock (Mach disk) in the near field - the flow region close to the nozzle exit. In some configurations the presence of secondary expansions and recompressions can be observed beyond  the first Mach disk (Belan et al. 2004, 2006).
%\cite{bel3,bel4}
In this experiment, the Mach number of the jets upstream of the disk is very high; values of up to 30 can be obtained. The Knundsen number $K$ always remains  in the continuum regime, except for particular cases (when the jet gas is helium with a 
high $p_0/p_{\rm amb}$ ratio, $K$ may approach the value of 0.5 in a small region upstream of the Mach disk). 

The time scale of the flow is defined as the time necessary for a fluid particle to cover the length of the vacuum chamber, which is of the order of 1 ms. The system permits an outflow of hundreds of time scales (typically 0.5 s). Since the primary pump has a very large delivery, we can assume a quasi-steady jet evolution. 

\begin{figure}[ht]
\centering
\includegraphics[width=12cm]{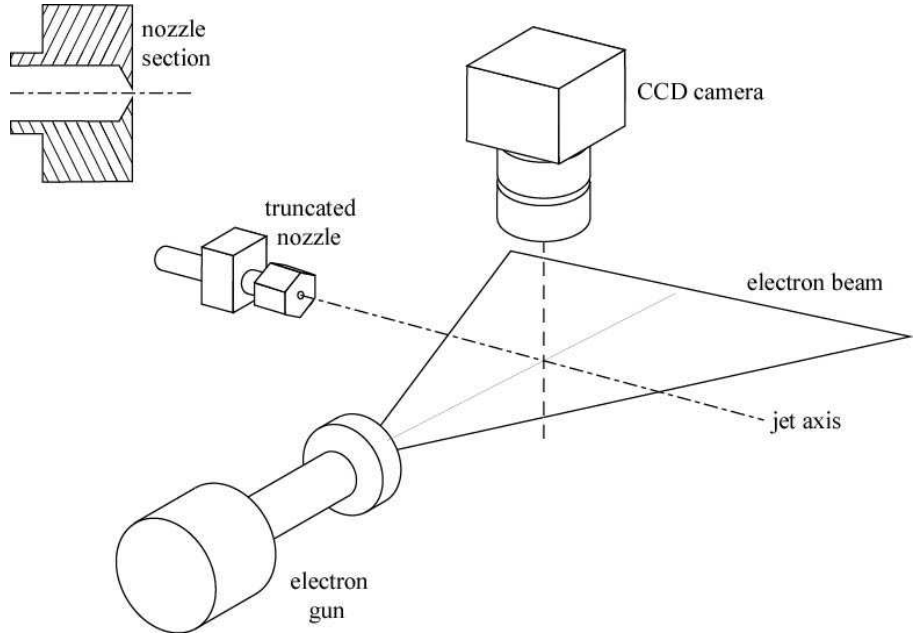}
\caption{Experimental setup. The truncated convergent nozzle is sonic at the exit section. The nozzle is here shown in the upper left corner.}
\label{setup}
\end{figure}

The experimental setup is shown in fig. \ref{setup}. The electron gun is operated at 16kV, with currents of  up to 2mA,  and is equipped with a deflection system that creates an electron blade. The blade intercepts the jet, and generates a plane
 fluorescent section of the flow, which is then acquired by an high sensitivity camera (1Mpixel CCD with Bayer RGB filter). 

This arrangement makes it possible to visualize the jet over many spatial scales, usually up to 200 initial diameters.
An important feature of the experiment is the possibility of studying  the effect of two main flow control parameters (the ambient/jet density ratio, and the Mach number) that can be set independently of each other.
Through the visualization of slices of the flow, and through the determination of the relevant density and species concentration distribution, it is possible to obtain information about the mixing layer  and the thickness of the shocks.

The present technique is subject to the conditions of proper working of a
continuous electron gun in a gas, that has a numerical density approximately
no greater than $10^{23}$m$^{-3}$ (for example, the best working range for air at
ambient temperature $T=300 K$ is $n < 2 \cdot 10^{22}\,$m$^{-3}$, which is equivalent to a mass density $\rho < 2\,$g/m$^3$, or a pressure $p < 300\,$Pa).

\section{A new procedure for the determination of density and concentration spatial distribution from fluorescent images}\label{concMisc}
The present method is based on the fluorescent emission from a gas excited by an electron beam. The relation between the radiation intensity $I$ and the numerical density $n$ of a  gas  (Brown and Miller, 1957) is
%\cite{brown}: 
% 
\beq 
  I \ =\ \frac{k\,\,n}{1 \,+\, h\, n}
%  I \ =\ \frac{k\,\,n}{1 \,+\, 2\, n\, \sigma^2 P_{jk}^{-1} \, \sqrt{4 \pi R_{gas} T}}
\label{grun}
\eeq
where $k$ is a constant that includes the sensitivity of the measuring system, and  $h$ is  a specific
coefficient  of the nature of the gas which depends on the temperature ($h = 2\, n\, \sigma^2 P_{jk}^{-1} \, \sqrt{4 \pi R_{gas} T}$, where $\sigma$ is the quenching collision diameter, $P_{mn}$ the spontaneous transitions
probability, $T$ the temperature and $R_{gas}$ the gas constant). Fluorescence occurs when the molecules excited by the electrons return to the ground state and
give back energy in the form of emitted light. The energy of incident electrons can also be transferred to translational molecular degrees of freedom or other molecular excited states (quenching collisions). In relation \rf{grun}, the $k n$ factor accounts for the proper fluorescence phenomenon, whilst the term $h n$ in the denominator accounts for the loss of fluorescence caused by quenching. 

\begin{figure}[ht]
\centering
\includegraphics[width=12cm]{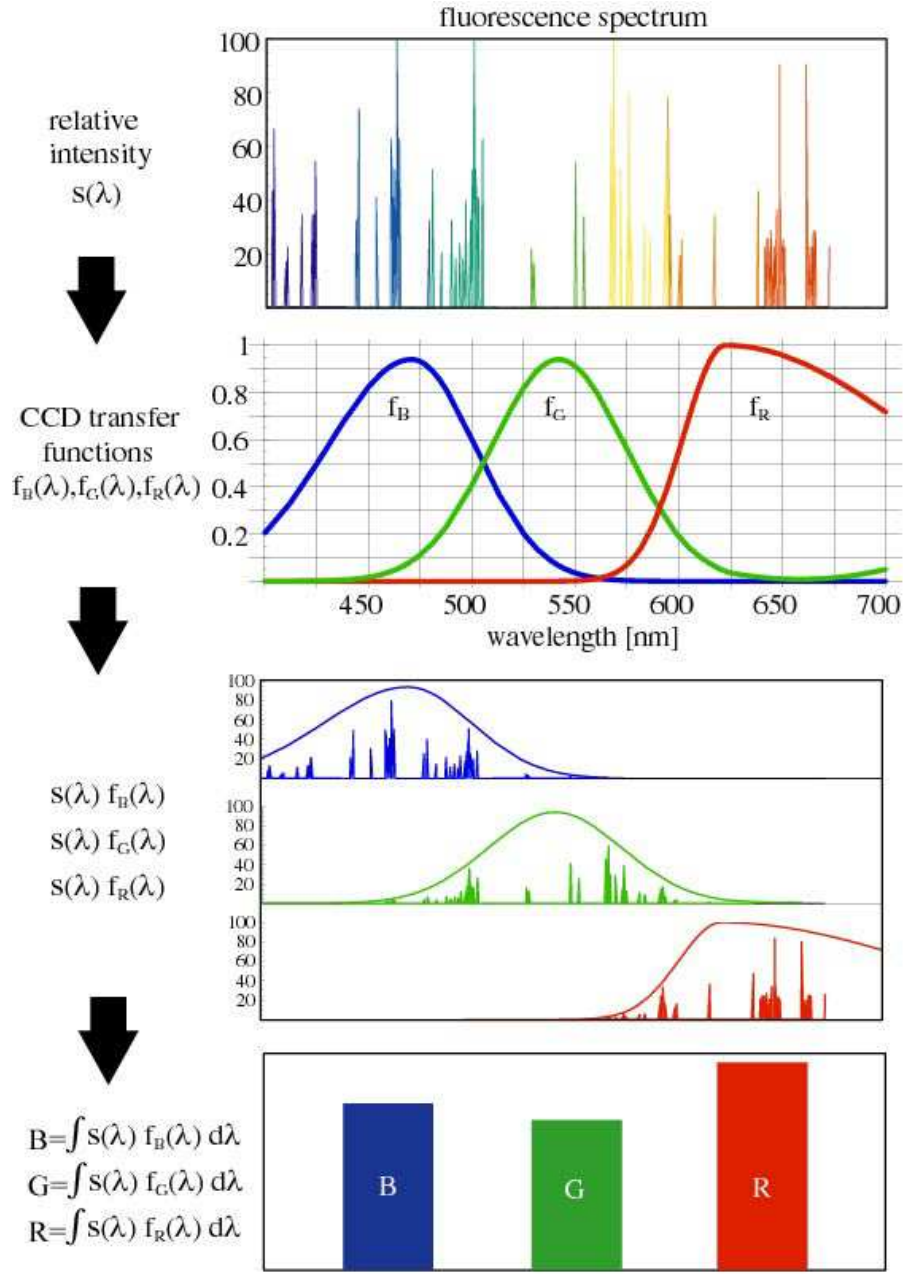}
\caption{Fluorescent emission acquired by a RGB sensor: the spectral lines are projected onto  the RGB response curves
of the CCD. The plots are normalized to maximum values.}
\label{spRGB}
\end{figure}

For relatively rarefied gases, that is, those in the small numerical density limit,  equation \rf{grun} shows that $I$ is proportional to $n$, whilst for higher densities, the 
emission encounters an asymptotic limit (saturation):
\beq
& &   I \, \propto \, n   \quad \textrm{as} \quad   n \to 0     \\ \label{Grun1}
& &   \lim_{n\to\infty} \!I \:=\: const. \qquad\quad\ \           \label{Grun2}
\eeq
At the low pressures considered here ($1.5 < p < 200$ Pa), we can approximate  equation \rf{grun} as 
\beq
I=k\, n ,
\eeq
Since $I$ is a spectral superposition, this relation also holds  for the three colours acquired and stored 
in the digital image yielded by the camera.  Three constants depending on the spectral emission range associated to each colour must then be introduced. In this study, we adopt the well-known RGB codification (R=red, G=green, B=blue, see fig. \ref{spRGB}). Thus, it is possible to write
\beq
R &=& k_R n \\
G &=& k_G n \\
B &=& k_B n ,
\eeq
where the constants $k_R, k_G, k_B$ are integral quantities that can be computed from known spectra or directly measured. It should be noted that for any linear superposition of  fluorescent emissions, 
\beq
C = a R + b G + c B,
\eeq 
one obtains:
\beq
C =  (a k_R + b k_G + c k_B) n = k_C n.
\label{clin}
\eeq
This relation also holds  for a two-gas mixture. If the mixture is sufficiently rarefied  to consider the radiation emission of the two gases as decoupled,  the total intensity will be the sum of the single ones. The present method assumes that this condition is satisfied.

If the ambient gas is labelled with $amb$ and the jet gas with $jet$,  we can write that the fluorescence emission $C$ is the sum of two expressions of the kind \rf{clin}:
\beq
C = C_{amb} + C_{jet} = k_{amb} n_{amb} + k_{jet} n_{jet} 
\label{ctot}
\eeq
By using again eq. \rf{clin} we obtain
\beq
k_C n = k_{amb} n_{amb} + k_{jet} n_{jet} 
\label{kc}
\eeq
By dividing by the total numerical density $n$ and introducing the concentrations $z_{amb}=n_{amb}/n$ for the ambient gas and $z_{jet}=n_{jet}/n$ for jet gas, the following  is obtained
\beq
k_C = k_{amb} z_{amb} + k_{jet} z_{jet}, 
\label{kz}
\eeq
where
\beq
k_a &=& a k_{Ramb} + b k_{Gamb} + c k_{Bamb},\label{KA}\\
k_j &=& a k_{R{jet}} + b k_{G{jet}} + c k_{Bjet}\label{KJ}.
\eeq
It should be noticed that the ratio between a couple of colour intensities takes different values for different gases. For example, the electron beam ionization at 16kV of pure helium in the pressure range 0.1 Pa$<p<$200 Pa at 300K gives an outstanding spectral line at 501nm, whilst in the same conditions pure argon gives a wide spectrum with several lines of comparable intensity, particularly in the red and blue zones. In this case, for example, the ratio $R/G$ will take on low values for helium and high values for argon.

Two linear superpositions of the kind
\beq
   C_1=a_1 R + b_1 G + c_1 B  \label{C1}  \\
   C_2=a_2 R + b_2 G + c_2 B  \label{C2}
\eeq
will usually give a ratio $C_1/C_2$ that varies with the species concentration. A suitable choice of the 6 coefficients
$a_i,b_i,c_i$ ($i$ = 1,2) can be made in order to obtain  a ratio  $C_1/C_2$ that varies over the largest possible interval for a given couple of gases, which in turn leads to the concentration determination.
   
In order to obtain a relation between the ratio $C_1/C_2$ and the concentration $z$ of a gas species
we choose a couple of different superpositions, and write their ratio using equations 
\rf{clin} and \rf{kz}:
\beq 
   \frac{C_1}{C_2}  = \frac{k_{C1}}{k_{C2}} = \frac{k_{amb_1} z_a + k_{jet_1} z_{jet}} {k_{amb_2} z_a + k_{jet_2} z_{jet} }.
\label{rel1}
\eeq
The concentrations of the two gases are linked by the relationships $z_{amb}=1-z_{amb}$.  Imposing $C_1/C_2 = r$,  equation \rf{rel1} becomes 
\beq \label{rel2}
r = \frac{k_{amb_1} \left(1-z_{jet}\right) + k_{jet_1} z_{jet}} {k_{amb_2} \left(1-z_{jet}\right) + k_{jet_2} z_{jet} }.
\eeq
By solving with respect to $z_{jet}$, one obtains
\beq \label{rel-Zg1}
  z_{jet} = \frac{k_{amb_1} - k_{amb_2} r}{k_{amb_1} - k_{jet_1} + \left(k_{jet_2} - k_{amb_2}\right) r}. 
\eeq
This gives  the concentration in the jet gas, for each pixel of the image, as a function of the ratio $r$ of the two superpositions $C_1, C_2$, and as a function of the four constant $k_{amb1},k_{amb2},k_{jet1},k_{jet2}$.
These constants are defined through (\rf{KA}-\rf{KJ}) and can be determined when the constants $k_{Ramb}$, $k_{Gamb}$, $k_{Bamb}$, $k_{Rjet}$, $k_{Gjet}$ and $k_{Bjet}$ are known and the six coefficients $a_i,b_i,c_i, \, i = 1,2$ are set for a given couple of gases. In synthesis, the $r$ variable is the ratio of the emission intensities measured in each pixel of the image for the two selected superpositions.

The $k$-constants in laws of the kind $I = k n$ are obtained from experimental calibrations in a known gas. By filling the vacuum vessel with a pure gas and keeping it at rest, with stable pressure $p$ and temperature $T$, it is possible to read a couple of values $I,p$. The repetition of this procedure with various pressures $p_i$ at the same $T$ gives an array of couples $\{I_i,p_i\}$. The perfect gas law  yields the numerical density $n=p {\cal N}_A/({\cal R} T)$ (where ${\cal N}_A$ and $\cal R$ are the Avogadro number and the universal gas constant, respectively), thus 
an array of values $\{I_i,n_i\}$ is obtained, see figure 4. $k$-constants are then deduced through a linear fit.  

\begin{figure}[ht]
\centering
\includegraphics[width=12cm]{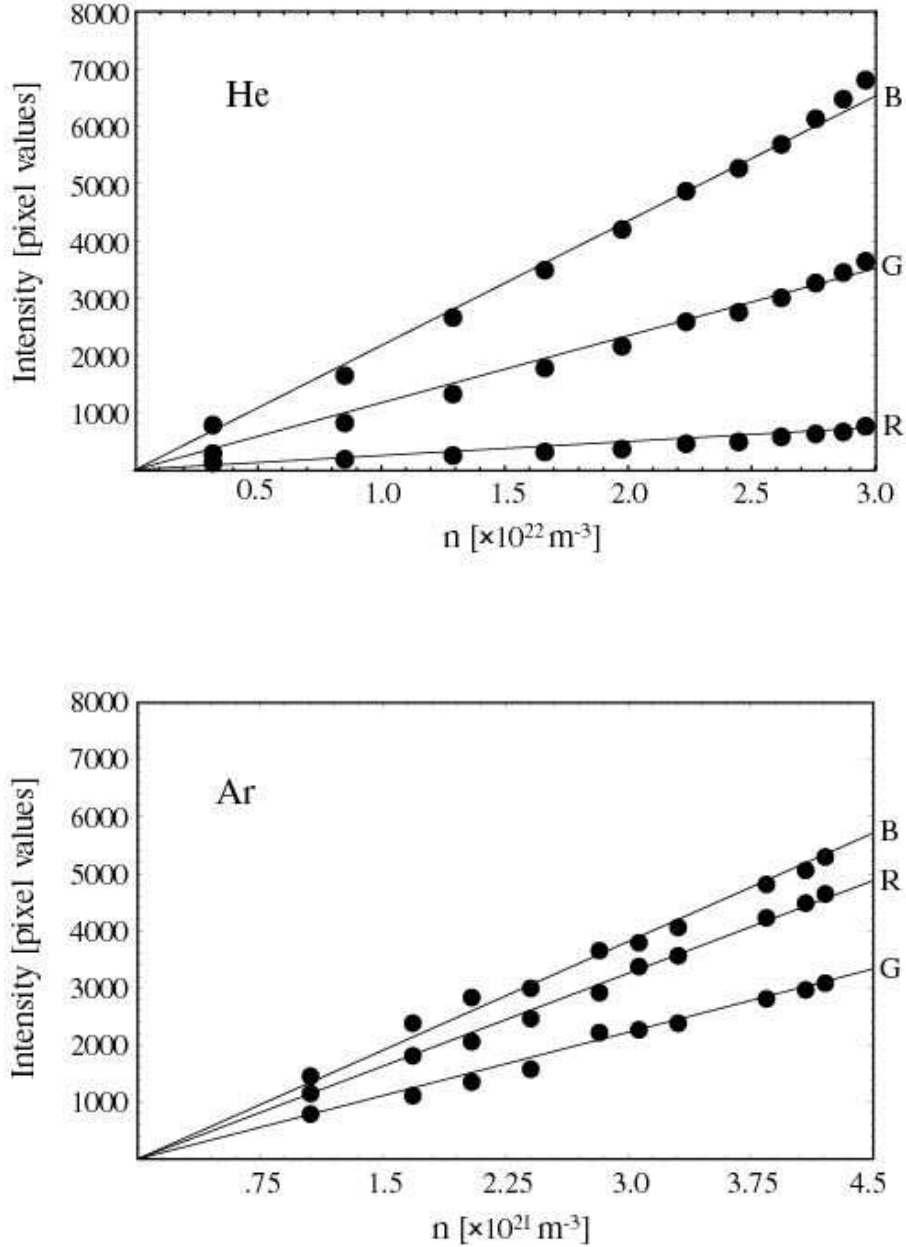}
\caption{Fluorescent emission acquired using an RGB sensor and varying the numerical density of the gas. The slopes of these curves are the constants $k_R, k_G, k_B$ in equations (5-7). 
Part (a): Helium calibration, variance values: 15\% for $k_R$, 12\% for $k_G$ and 6\% for $k_B$. 
Part (b): Argon calibration, variance values:  3.5\% for $k_R$, 6.8 for $k_G$ and 4.9\% for $k_B$}
\label{calib}
\end{figure}

The four constants $k_{amb_1}$, $k_{amb_2}$, $k_{jet_1}$, $k_{jet_2}$ are not independent, since, by means of  equation \rf{rel1}, it is possible to write the following for the pure ambient gas ($z_{amb}=1, z_{jet}=0$) and for the pure jet gas ($z_{amb}=0, z_{jet}=1$)
\beq   
   \frac{C_{amb_1}}{C_{amb_2}} = \frac{k_{amb_1}}{k_{amb_2}} \;\;;\;\;
   \frac{C_{jet_1}}{C_{jet_2}} = \frac{k_{jet_1}}{k_{jet_2}}
\eeq
so that
\beq   
   k_{amb_1}= k_{amb_2} \frac{C_{amb_1}}{C_{amb_2}} \;\;;\;\;  k_{jet_1}= k_{jet_2} \frac{C_{jet_1}}{C_{jet_2}}. \,
\eeq
Equation \rf{rel-Zg1} can thus be written in the form
\beq \label{iperb-z}
  z_{jet} = \frac{A - r}{\lp(A - r\rp) - M\lp(J - r\rp)}
\eeq
where $r$ is the independent variable (obtained from each image pixel, using the chosen set of coefficients 
$a_1, b_1, c_1$ and $a_2, b_2, c_2$), 
$A = {C_{amb_1}}/{C_{amb_2}}$ is a known constant for the pure ambient gas, 
$J = {C_{jet_1}}/{C_{jet_2}}$ is a known constant for the pure jet gas,
$M = {k_{jet_2}}/{k_{amb_2}}$ is a ratio of known constants that are separately determined and are associated to the triplet $a_2,b_2,c_2$ ($k_{jet_2} =  a_2 k_{Rjet} + b_2 k_{Gjet} + c_2 k_{Bjet}$ and $k_{amb_2} =  a_2 k_{Ramb} + b_2 k_{Gamb} + c_2 k_{Bamb}$). The constants $A,J$ and $M$ are calculated from the known $k$-constants.
Equation \rf{iperb-z} is a hyperbolic law which is physically meaningful for values of $z_{jet}$ in the range 
$0 \le z_{jet}\le 1$, corresponding to a range $r_{min}< r < r_{max}$.

Once the concentrations are known, the density can be determined, as the total numerical density $n$ may be obtained from relations such as  (\ref{clin}), e.g. one can use 

\beq
C_1 =  k_{C_1} n,  \;\;\; k_{C_1}=  k_{amb1} z_{amb} + k_{jet1} z_{jet}
\label{1clin}
\eeq

or

\beq
C_2 =  k_{C_2} n, \;\;\;  k_{C_2}=  k_{amb2} z_{amb} + k_{jet2} z_{jet}.
\label{2clin}
\eeq

The density is finally given by the equation
\beq
\rho = n (z_{amb} m_{amb} + z_{jet} m_{jet}) 
\label{DENS}
\eeq
where $m_{amb}, m_{jet}$ are the molecular masses of the gases in the ambient and in the jet, respectively.

It is worth noting that since the values of $r=C_1/C_2$ are obtained from the image analysis, the image noise
may lead to $z_{jet}$ values that lie outside the physical range. The  noise is particularly dangerous when it induces
a value $r$ close to the value $r_s =(MJ-A)/(M-1)$ that is a singularity of equation \rf{iperb-z}, 
which however lies outside the physical domain $r_{min}< r < r_{max}$. 

\begin{figure}[ht]
\centering
\includegraphics[width=12cm]{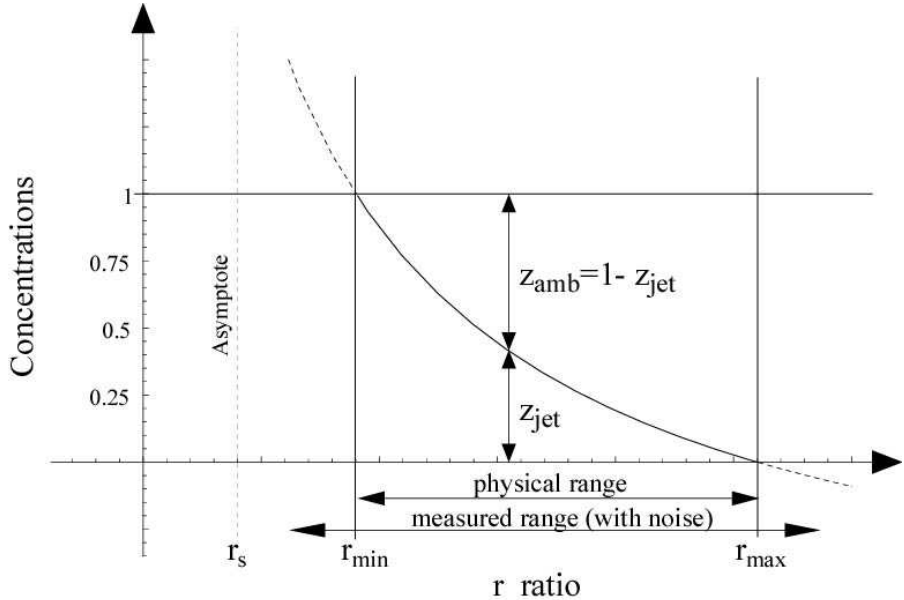}
\caption{Typical dependence of the concentration z on the color ratio r. The measured range of the variable $r$ is usually larger than the physical range because of noise.}
\label{PLOTXZ}
\end{figure}

The noise effects are sketched in fig. \ref{PLOTXZ}. It should be noted that it is advisable to adopt a noise filtering technique in the image post-processing. 
In particular, the filter must preserve the primary ratios among the three colors R, G and B along the image.

It should also be noticed that this procedure is not working when the fluorescent spectra of the ambient gas  and of the jet gas are similar (for example air and Argon). In this case, constants $A$ and $J$  become nearly equal and constant $M$ becomes nearly equal to 1 and as a consequence  the concentration measurement becomes impossible in this case.

The present measurements are always performed on 1300x1030 RGB images, with a 10 bit resolution. Images were obtained by operating the electron gun at 16kV with a beam current of 1.5 mA.

\section{Results and discussion}

The procedure was applied to measure the concentration $z$ and the density $\rho$ in two underexpanded jets. The first is an under-dense jet of Helium in an Argon ambient and the relevant pressure ratio is $0.84 \cdot 10^3$. The second is an over-dense jet of Argon in a Helium ambient where the pressure ratio is $1.2 \cdot 10^3$. The exposure time was 1/12 s, 
much longer than the flow time scale (about 1 ms), but shorter then the outflow time, which is  $\sim 0.5$ s. Thus, each image can be assumed to visualize a flow in quasi-steady conditions. 
The images are compensated for the non-uniformity of the electron blade, and filtered to improve the signal-to-noise ratio by means of local averages in small neighborhoods of each pixel.

%%%%%%%%%%% He in Ar

As a first example, we considered the  Helium jet flowing in an argon ambient. The  choice of the coefficients was $a_1=1, b_1=1,  c_1=1$; $a_2=0, b_2=1,  c_2=0$. The calibration led to:
  
$A = {C_{a1}}/{C_{a2}} = 4.18 \pm 0.1$  for pure Ar (ambient gas), 

$J = {C_{j1}}/{C_{j2}} = 3.06 \pm 0.1$  for pure He (jet gas),

$M = {k_{j2}}/{k_{a2}} = 1.58 \pm 0.06$  for the He-Ar couple  (jet/ambient)

The physical range for $r$, see fig. \ref{PLOTXZ}, is in this case $3.06<r<4.18$.

\begin{figure}[ht]
\centering
\includegraphics[width=12cm]{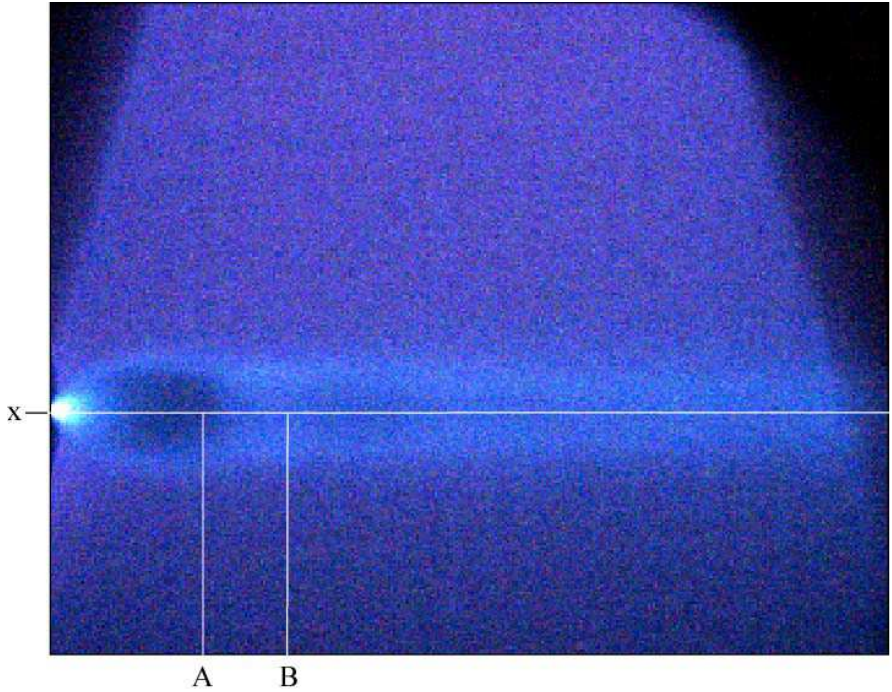}
\caption{Helium jet in an Argon medium: pressure ratio $p_0/p_a \sim 0.84 \cdot 10^3$,  
Mach before the normal shock $\sim 26$, Reynolds number at the nozzle exit =3000.  
The straight lines (A,B) mark the measurement sections. 
}
\label{HEAR}
\end{figure}
Figure \ref{HEAR} shows a Helium jet obtained with a stagnation pressure $p_0=(3.06 \pm 0.20)\cdot 10^4$ Pa, whilst 
the  pressure of the argon ambient is $p_a=36.5\pm 0.5$ Pa. The figure is strongly contrasted for clarity. 
The measurements are performed along the $x$ axis and along two cross-sections, section A before the Mach disk and section B after the disk. 
It should be noticed that  the Knudsen number upstream of the Mach disk in this jet is quite high, about 0.5. This explains why the shock is very thick. 

\begin{figure}[ht]
\centering
\includegraphics[width=12cm]{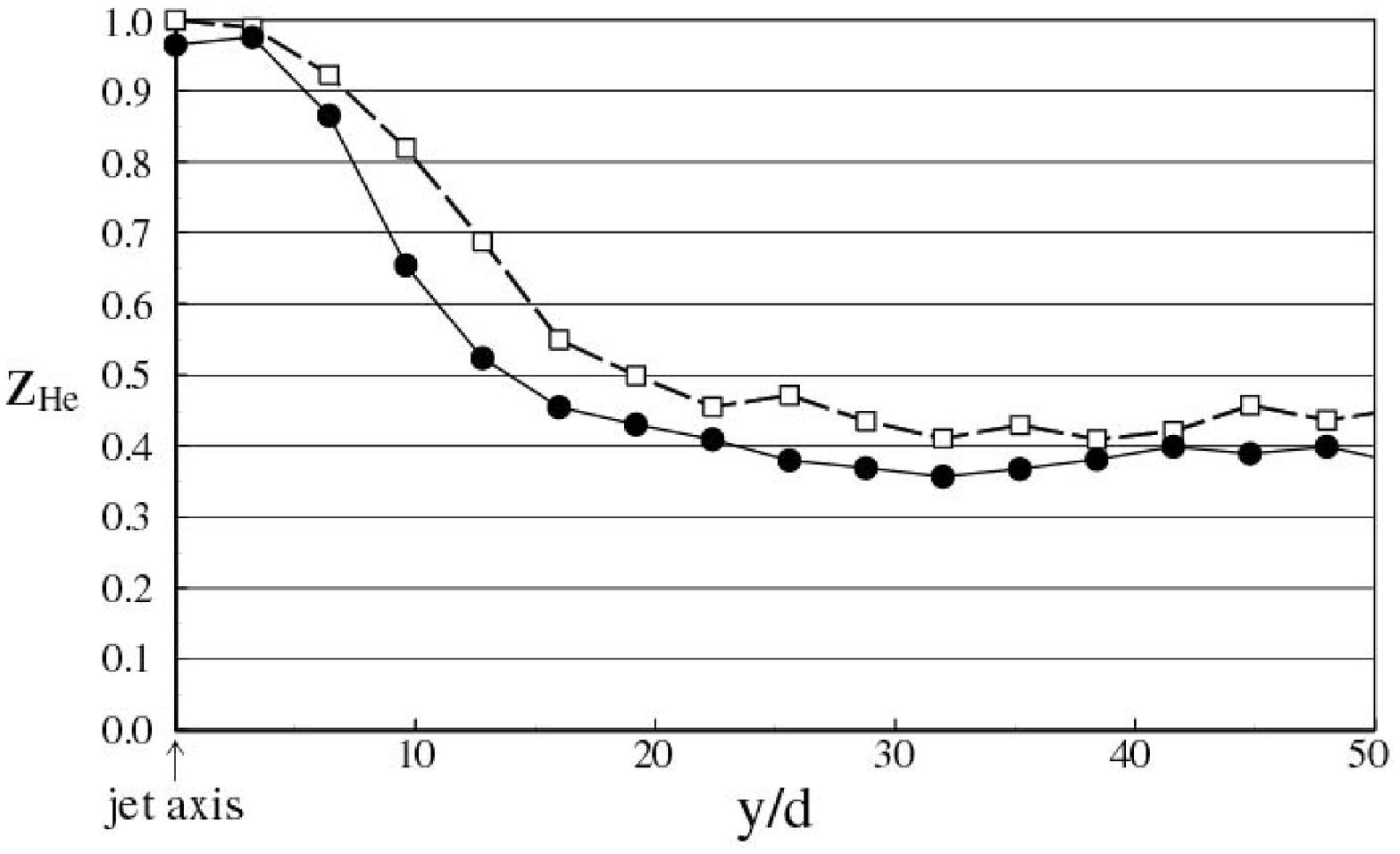}
\centering
\includegraphics[width=12cm]{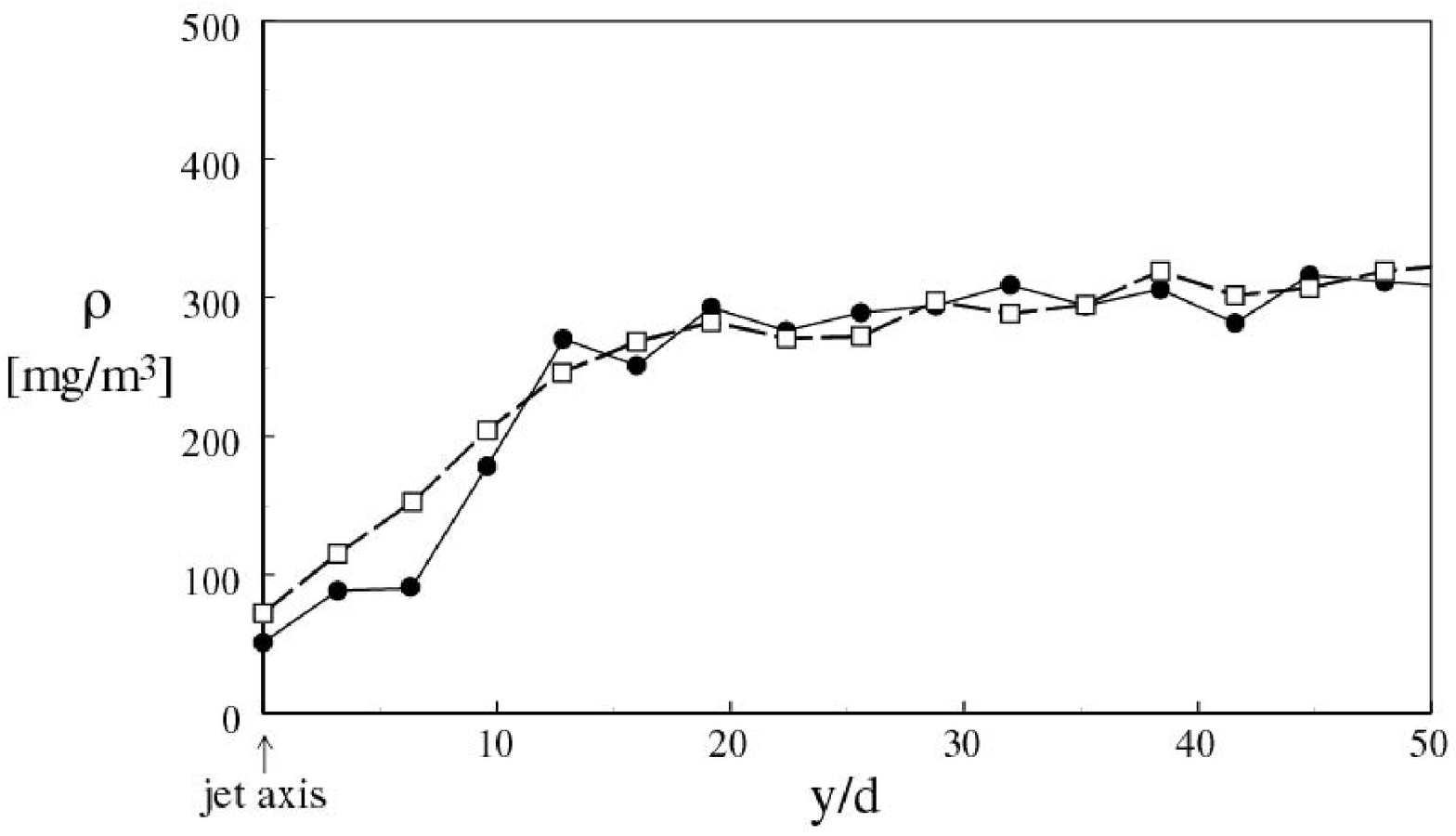}
\caption{He jet in an Ar ambient, see visualization in  fig. \ref{HEAR}, $p_0/p_a = 0.84 \cdot 10^3$, 
Mach before the normal shock $\sim 26$, nozzle exit Reynolds number =3000. 
Part (a): Cross-sectional concentration. 
Part (b): Cross-sectional  density.
$\bullet$: section (A), before the Mach disk. 
$\Box$: section (B), after the Mach disk.}
\label{zyHEAR}
\end{figure}

Figure \ref{zyHEAR}a shows the concentration curves of this jet at the cross-sections A ($x/d=17.6\pm0.5$) and B 
($x/d=26.4\pm0.5$). The radial distance is adimensionalized over the initial jet diameter $d$=2mm, which is the diameter of the nozzle orifice. The highest helium concentration is on the jet axis and the lower concentrations are in the outer zone. In particular, the curves show that  the mixing with the ambient is more effective after the shock, since the helium concentration increases after the Mach disk.

Figure \ref{zyHEAR}b shows the density distributions. 
This is the total density obtained from equation \rf{DENS}. It can be  seen that the jet remains underdense with respect to the surrounding ambient even in the post-shock zone. 

\begin{figure}[ht]
\centering
\includegraphics[width=12cm]{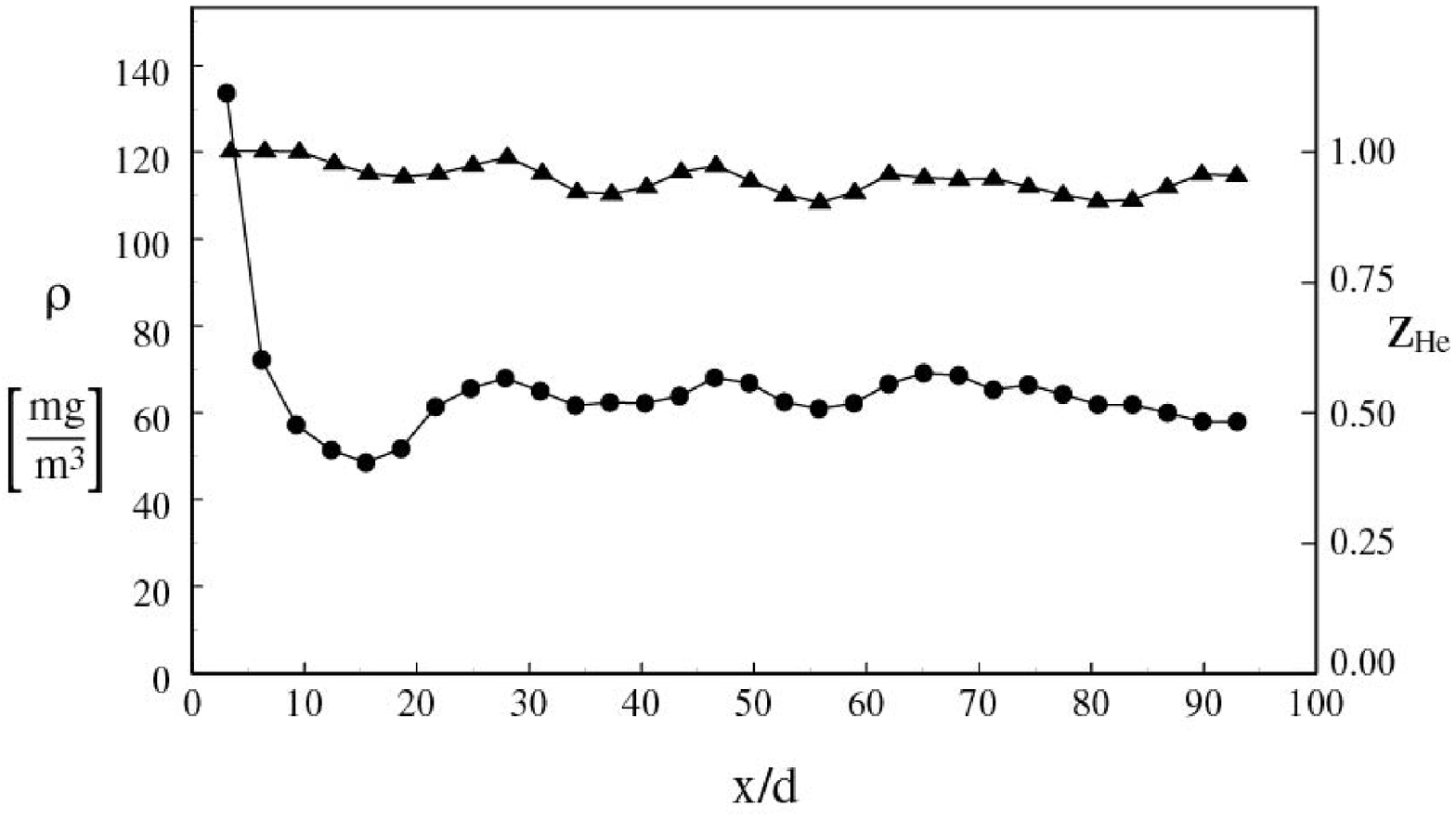}
\caption{Axial concentration and density for the Helium jet visualized in fig. \ref{HEAR}, 
$p_0/p_a = 0.84 \cdot 10^3$,  Mach before normal shock  $\sim 26$, Reynolds at the nozzle exit =3000. 
$\bullet$: density, $\triangle$: concentration.
}
\label{roxHEAR}
\end{figure}

Figure \ref{roxHEAR} shows the axial density and He concentration in the jet (the first point on the left of the curve is omitted because of image saturation). It can be seen that the shock thickness is remarkable. In this jet, it is also possible to see secondary expansions and compressions after the Mach disk. The shock center location 
is estimated as
$x_M = (21.7 \pm 0.5) d$, which is in good agreement with the theoretical shock position given by Young (1975)
\beq
x_M = C_\gamma d \sqrt{p_0/p_a}
\label{YOUNG}
\eeq
where $C_\gamma$ is a  constant that depends on the specific heat ratio of the gas in the jet, in this case $C_\gamma = 0.76$ and $p_0/p_a = 0.84 \cdot 10^3$, which gives $x_M/d=22.01$. The shock thickness $\Delta$ is of the order of $11 d$. The  concentration of Helium is equal to 1 at the nozzle exit and remains almost constant along the axis. Mild oscillations in phase with that of the density are observed downstream of the Mach disk.

According to Ashkenas et al. (1966) and Young (1975), the maximum Mach number before the shock can be estimated  as 
$M_{max} \simeq 26$ at $x=x_M$. This is obtained through a semi-empirical model which assumes that the rapid expansion downstream of the nozzle exit is isentropic and that the streamline in the region of the flow inside the barrel shock satisfies a spherical symmetry  close to the axis.

%%%%%%%%%%% Ar in He

The second test case we present  is an Argon jet flowing  in a Helium ambient. This jet could be considered as the complementary configuration of the previous case. The exchange of gases in the jet and in the ambient produces an over-dense jet in  place of the precedent under-dense case. The choice for the 6 coefficients $a_i, b_i, c_i, i = 1,2$ of the algorithm  is the same as that of the  previous test case. The  ambient gas/jet gas exchange yields

$A =3.06 \pm 0.1$ 

$J =4.18 \pm 0.1$ 
 
$M =0.63 \pm 0.06.$

\begin{figure}[ht]
\centering
\includegraphics[width=12cm]{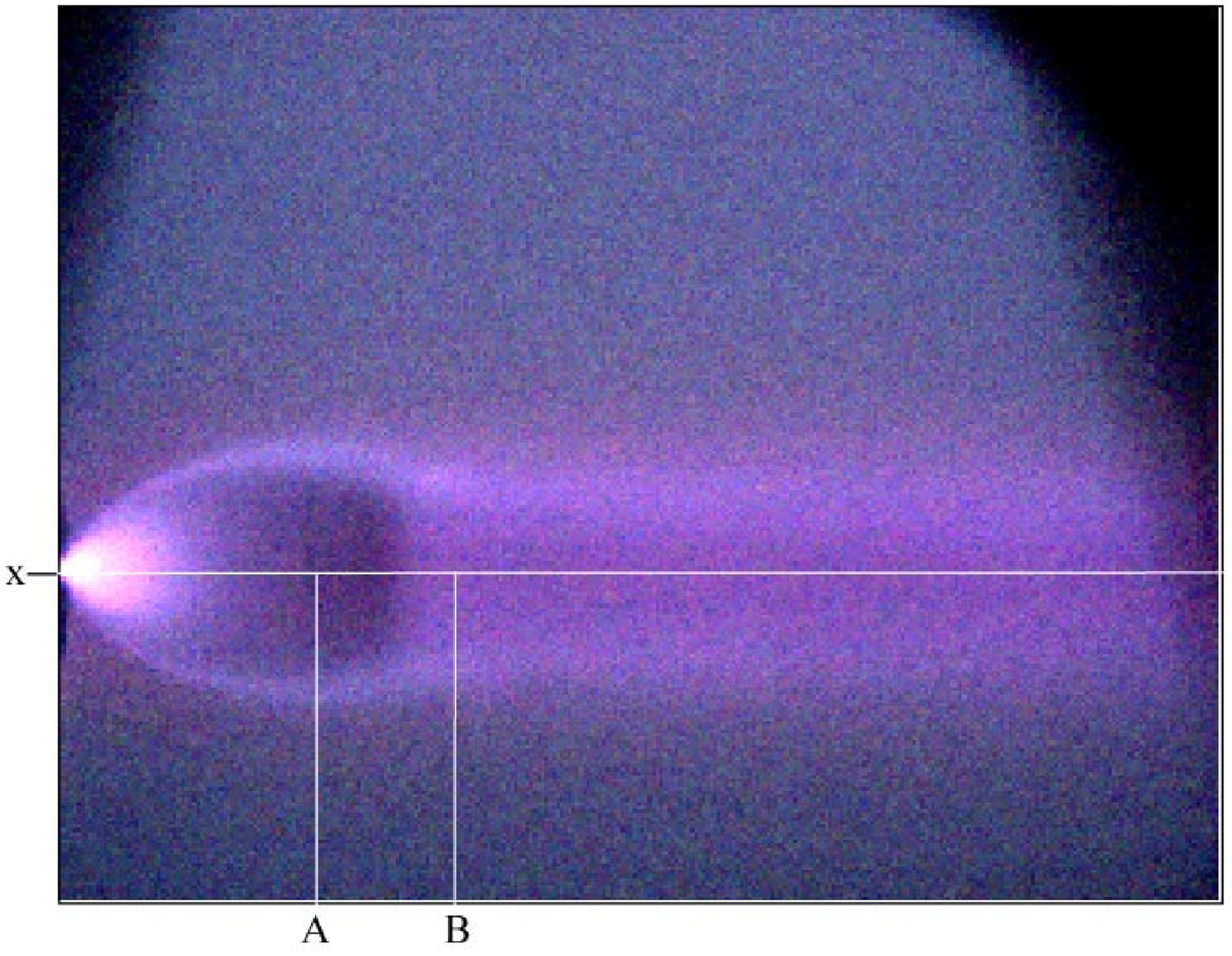}
\caption{Argon jet in a Helium medium: pressure ratio $p_0/p_a = 1.2 \cdot 10^3$, Mach before the normal shock $\sim 29$, Reynolds at the nozzle exit =18200.  The straight lines (A,B) indicate the cross-stream measurement sections.}
\label{ARHE}
\end{figure}
Figure \ref{ARHE} shows a visualization  image of the Argon jet, which is flowing in similar conditions to those applied to the complementary jet visualized in fig. \ref{HEAR}, since the stagnation/ambient pressure ratio is of the same order ($p_0/p_a = 1.2\cdot 10^3$). 
Here the stagnation pressure is $p_0=(5.68 \pm 0.20) \cdot 10^4$ Pa, and the helium ambient pressure is 
$p_a=(45.7\pm0.5)$ Pa. As in the previous case, the measurements are performed along the $x$ axis and along two cross-sections A and B.

\begin{figure}[ht]
\centering
\includegraphics[width=12cm]{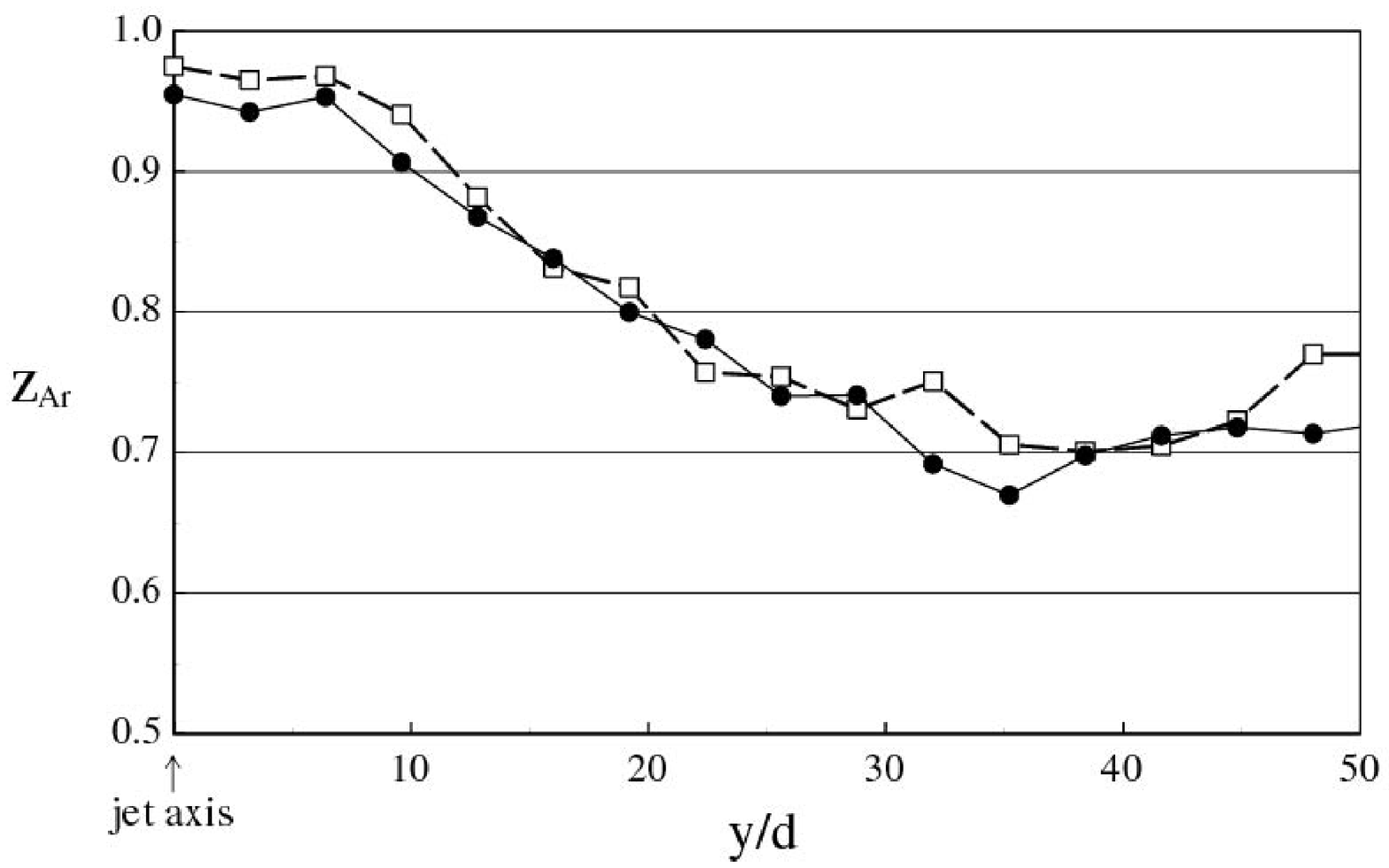}
\centering
\includegraphics[width=12cm]{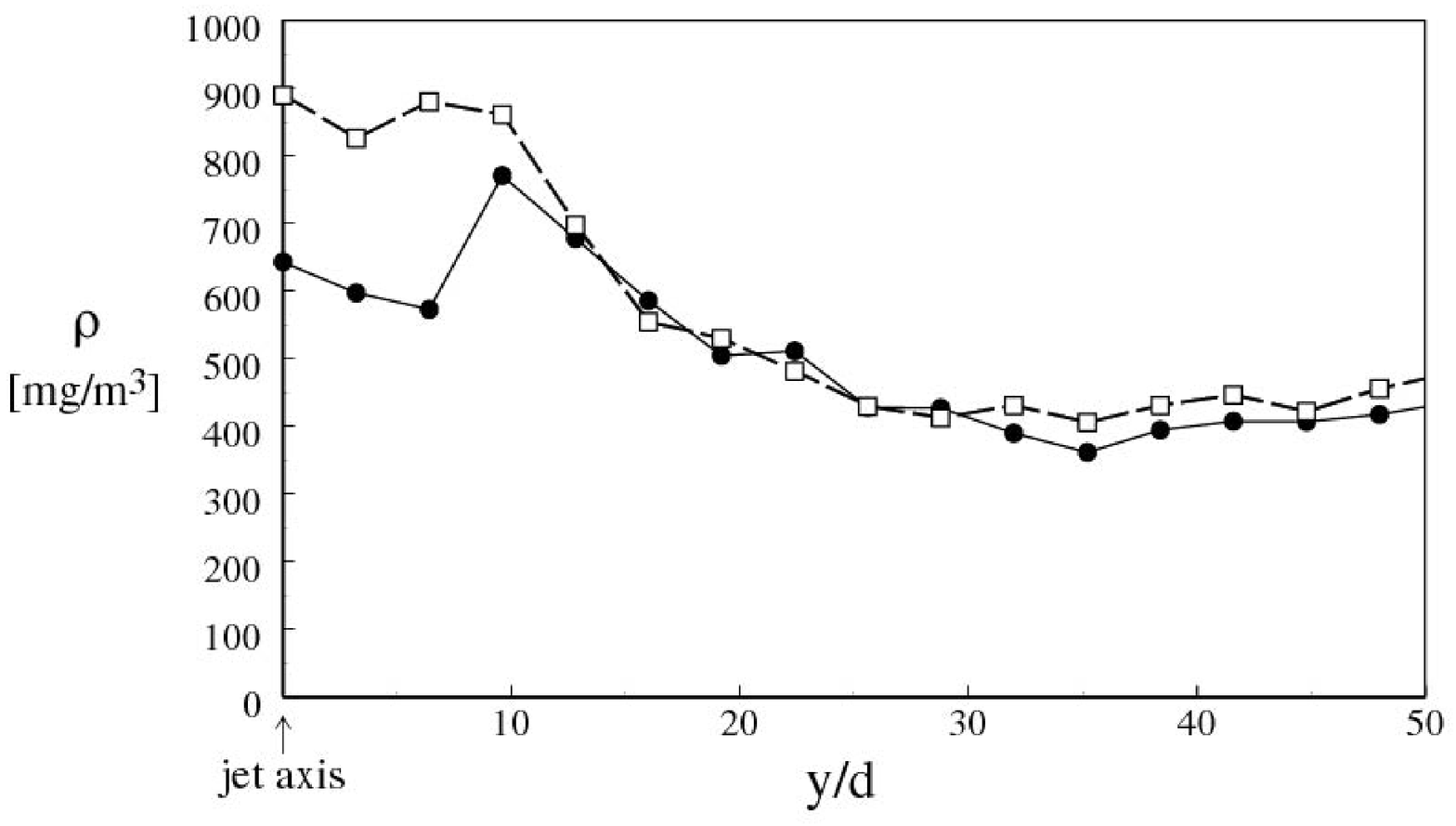} 
\caption {Argon  jet in a Helium ambient, see visualization in fig. \ref{ARHE}, $p_0/p_a = 1.2 \cdot 10^3$,  Mach before the normal shock  $\sim 29$, Reynolds at the nozzle exit =18200. Part (a): cross-sectional concentration. Part (b): cross-sectional  density. 
$\bullet$: before the Mach disk. $\Box$: after the Mach disk.}
\label{zyARHE}
\end{figure}

Figure \ref{zyARHE}a shows the concentration curves of this jet at the cross-sections A ($x/d=20.5\pm0.5$) and B 
($x/d=30.7\pm0.5$). In this case, the diffusion of the gas, the heavy Argon, in the surrounding ambient is very effective. In fact, the argon concentration in the ambient is  larger than 0.7 throughout. In the Helium jet case, the lateral far field concentration of Helium was 0.4. The lateral spreading of the jet is also larger, and is of the order of 35-40 nozzle exit diameters compared to  the 20 diameters found for the Helium jet. 
Figure \ref{zyARHE}b shows two cross-sectional density curves of the same jet, as in figure \ref{zyARHE}a. Here the jet is
always overdense and  remarkably so in the mixing layer zone. 

\begin{figure}[ht]
\centering
\includegraphics[width=12cm]{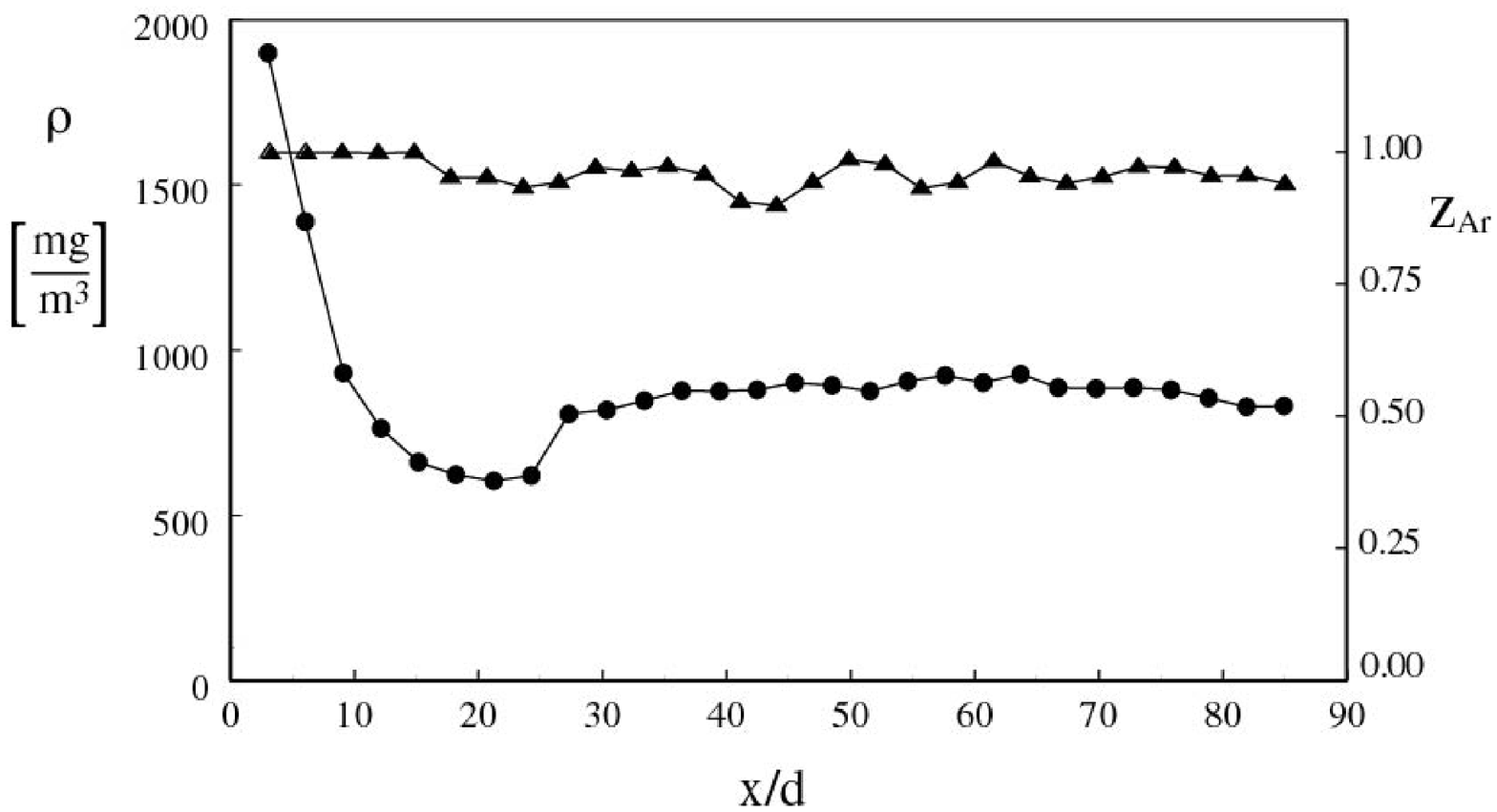}
\caption{Axial concentration and density in the jet, see visualization in fig. \ref{ARHE}, $p_0/p_a = 1.2 \cdot 10^3$,  Mach before the normal shock  $\sim 29$, Reynolds at the nozzle exit =18200. $\bullet$: density; $\triangle$: concentration.}
\label{roxARHE}
\end{figure}

Figure \ref{roxARHE} shows the axial density and Ar concentration distributions in this jet. In this case, the shock is thinner ($<3d$), and there are no further expansions or compressions downstream. The estimated shock center location 
$x_M = (27 \pm 0.5) d$ is in good agreement with the prediction $x_M = 26.8 d$ given by equation \rf{YOUNG} with a pressure ratio $p_0/p_a = 1.2 \cdot 10^3$ (constant $C_\gamma$ is again equal to 0.76  because Ar and He are both mono-atomic). 
The estimate of the maximum Mach number upstream of the shock is  $29$. The concentration of the gas along the jet axis  in this case is again almost constant and mild oscillations are again visible
downstream of the Mach disk.

\begin{figure}[ht]
\centering
\includegraphics[width=12cm]{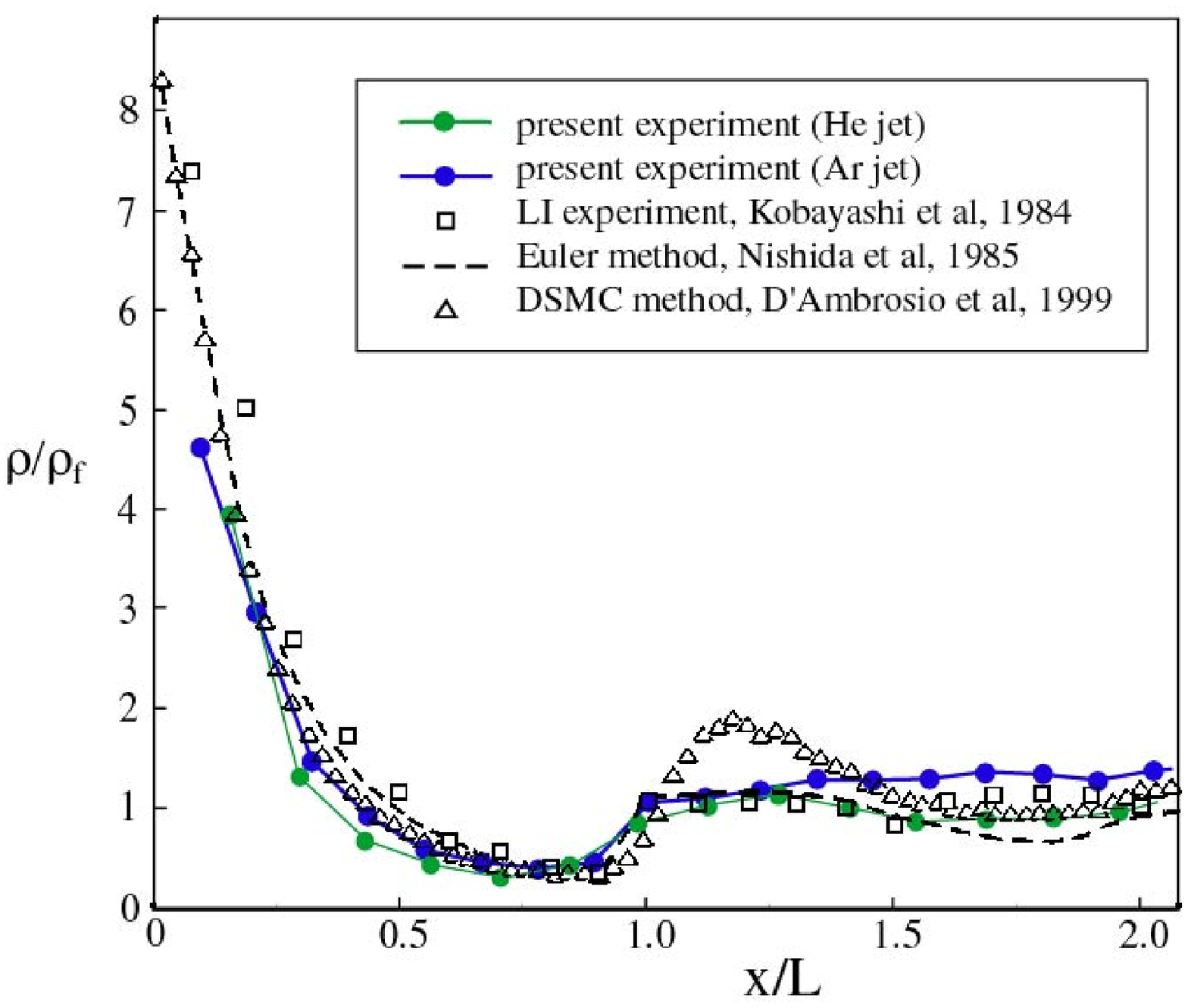}
\caption{Comparison of the streamwise axial density distribution in different under-expanded jets. Black $\bullet$: He in Ar,  $Re = 3000$,  $M =26$. Blue $\bullet$: Ar in He, $Re=18200, M = 29$. $\Box$: $Re=3800, M = 6.7$.  - - - :  Ar in Ar, $Re = 3800, M = 6.7$. $\triangle$: air in air, $ M = 5$. $M$ is the maximum Mach number in the jet which is encountered in front of the Mach disk, $Re$ is the Reynolds number at the nozzle sonic exit. }
\label{CONF}
\end{figure}

We conclude the discussion of the results by presenting, in figure \ref{CONF}, a comparison with results obtained by other researchers using very different 
methodologies, such as a numerical simulation of an Argon jet in an Argon ambient based on Euler's  equations (Nishida et al. 1985), a numerical simulation of an air jet in air  based  on the Monte Carlo direct simulation  method (D'Ambrosio et al. 1999), and  laboratory laser interferometer measurements of an Argon jet in an Argon ambient (Kobayashi et al. 1984).
The compared jets are characterized by different values of two control parameters, the Reynolds number (in particular the Re based on flow values at the nozzle sonic exit) and the Mach number reached before the normal shock. It can be noticed that the axial densities collapse very well on a single self-similar curve, which is obtained by scaling the longitudinal distance with the axial length of the barrel shock $L$ and by scaling the density with the value reached beyond the normal shock. This comparison shows that the present density determination procedure from fluorescent images is effective. From the dynamical point of view, this result has highlighted the Reynolds and Mach number similarity of the near and intermediate field of the under-expanded jet.

The overall accuracy of these measurements is about 10\%. The uncertainty grows from 5\% in the central part of the jet to  10\%  for the concentration and  15\% for the density  in the lateral far field. 
For the longitudinal distributions, the uncertainty grows from  5\% inside the barrel shock region to  10\% in the  far field downstream of the Mach disk.

\section{Conclusions}

We propose a new procedure to exploit the information included in the  fluorescent image of a gas in motion to determine the distribution of species concentration and density.
The fluorescent emission is produced by the excitation of atoms/molecules of the gas when an electron blade crosses it. 
The intensity of the fluorescent light is proportional to the local number density  of the gas. When the gas flow  is a mixture of  different species, 
this proportionality allows the contribution associated to different chemical species from the spectral superposition acquired by a digital camera to be determined. This yields a mean of simultaneously obtaining  species concentration and   mass density in a gas mixture flows.
The procedure was  applied to two underexpanded gas jets discharged into a different gas ambient - Helium into Argon and Argon into Helium. The obtained density distributions contrast satisfactory with other experimental and numerical determinations.

The limits of the present measurements are essentially linked to  the saturation of the emission intensity at  high  pressures. In the underexpanded jets here tested, the limit (in this case, pressures larger than $200$ Pa) only pertains  to a very small region close to the sonic nozzle, where  the pressure and density are very high and the density is underestimated. In this zone, the measured concentration may differ from the theoretical unitary value because of the saturation of the RGB colours. 

The results considered here show that this technique could  be useful to  measure mixing layer thicknesses, since the
mixing region between the jet and the surrounding ambient is in the best working range for this algorithm.
This suggests that the technique may be used in other gas flows with shear layers, provided  the fluorescent
spectra of the gases are sufficiently different. 

The present technique is not limited to applications pertaining to highly compressible flows. In fact, gas flows that are incompressible, but stratified in density and/or species concentration, can be considered provided the pressure remains below a threshold of about $200 \sim 300 $ Pa (depending on the gas nature).

A comparison with results obtained from other authors on under-expanded jets with different values of Mach and Reynolds numbers has resulted to be very satisfactory. One result associated to the validation analysis of the present procedure was the confirmation that the longitudinal evolution of an under-expanded jet becomes  almost self-similar when scaling the streamwise coordinate with the barrel shock length.

\end{document}